\documentclass[aps,nofootinbib,floatfix,showpacs,preprintnumbers,prd]{revtex4}
\usepackage{graphicx, epsfig, bm, amsmath}
\usepackage{amssymb}
\usepackage{color}
\usepackage{float}
\usepackage{hyperref}

\usepackage{wasysym}

\begin{document}


\title{Study of the very high energy gamma-ray spectrum from the Galactic Center and future prospects}

\author{Alexander V. Belikov$^{1,2}$, Emmanuel Moulin$^3$ and Joseph Silk$^{1,4,5}$}
\affiliation{$^1$Institut d'Astrophysique de Paris, UMR 7095 CNRS, Universit\'e Pierre et Marie Curie, 98 bis Boulevard Arago, Paris 75014, France}
\affiliation{$^2$Knowledge Lab, Computation Institute, University of Chicago, Chicago IL 60637, USA}
\affiliation{$^3$DRF/Irfu, Service de Physique des Particules, CEA Saclay, F-91191 Gif-Sur-Yvette Cedex, France}
\affiliation{$^4$Department of Physics and Astronomy, 3701 San Martin Drive, The Johns Hopkins University, Baltimore MD 21218, USA}
\affiliation{$^5$BIPAC, 1 Keble Road, University of Oxford, Oxford OX1 3RH UK}
\begin{abstract}
Ground-based gamma ray observations of the Galactic Center region have revealed a high energy gamma ray source spatially coincident with the gravitational centroid of our Galaxy. The pointlike emission detected by H.E.S.S. exhibits an energy cut-off at about 10 TeV.
We identify the parameters of the best fit of the exponential and the super-exponential cutoff models to the spectrum of the pointlike source and find that super-exponential one provides a similar quality of the fit to the spectrum of the pointlike source as the best-fit exponential cutoff model, while a dark matter interpretation does not provide as good a fit in the whole energy range of the signal.
Based on the magnitude of the flux we derive constraints in the plane of the slope of the density profile $\gamma$ and the critical radius, below which the density is assumed to be constant, $r_c$.
Motivated by recent results on the spectrum and morphology from H.E.S.S. and by the possible observation of a super-exponential cutoff, we forecast the observations of super-exponential versus exponential cut-offs by the upcoming Cherenkov Telescope Array (CTA). We derive a formula for $J$-factor in the small angle approximation and propose approximate morphological constraints on the central source.
\end{abstract}


\maketitle

\section{Introduction}
Major experimental efforts have been carried out in order to detect the non-gravitational interactions of dark matter (DM) in the past two decades. One of the premier strategies developed to unveil the identity of DM is based on the {\it indirect} detection through the hunt of the final state annihilation products such as gamma rays, cosmic rays and neutrinos. High energy gamma rays are powerful probes: since they do not suffer from propagation effects at the galactic scale, they can reveal the distribution of dark matter. Distinct features such as a steep spectral cut-off, lines, or bumps, may be present in the spectrum allowing for efficient discrimination against standard astrophysical emissions. Central regions of galaxies where dark matter signals are expected to be large are particularly promising regions to look for dark matter annihilations through gamma rays.

The H.E.S.S. array of ground-based Cherenkov telescopes in Namibia is well located for very-high-energy (VHE, E$>$100 GeV) observations of the Galactic Center region.
H.E.S.S. has accumulated more than 220 hours of observations at the nominal position of the Galactic Center since 2004 and provides the most detailed picture to date in VHE gamma-rays of the inner 200 pc of Galactic Center region~\cite{Abramowski::2016dhk}. A strong emission coincident in position with the supermassive black hole Sgr A* lying at the dynamical center of the Milky Way~\cite{Aharonian:2004wa,Albert:2005kh,Aharonian:2009zk,Archer:2014jka} has been detected. The position of the centroid of the HESS J1745-290 emission is coincident with Sgr A* and the pulsar wind nebula G359.95-0.04 positions within 13$^{''}$~\cite{Collaboration:2009tm}. In addition, a diffuse VHE emission extended along the Galactic plane in the inner 200 pc has been reported~\cite{Aharonian:2006au,Abramowski::2016dhk,2015arXiv151004518L}.
In the lower energy range 100 MeV-100 GeV the HESS source J1745-290 was identified as source 3FGL J1745.6-2859c by $Fermi$ LAT collaboration~\cite{Acero:2015hja}.

The central emission in the region of 15 pc ($\sim$ 0.1$^{\circ}$) radius, namely the pointlike VHE source HESS J1745-290, shows a significant deviation from a pure power law emission in a few TeV energy bands and the data~\cite{Aharonian:2009zk,VianaICRC,Abramowski::2016dhk} are well described by a power law with an exponential energy cut-off (EPL): $\Phi_{EPL} (E) = A_s E^{-\Gamma_s} e^{-E/E_c}$, or a power law with a super-exponential energy cut-off (SEPL): $\Phi_{\rm SEPL} (E) = A_s E^{-\Gamma_s} e^{-(E/E_c)^\beta}$, where $A_s$ is the normalization factor, $\Gamma_s$ is the exponent of the power law, $E_c$ is the energy cutoff and $\beta$ is the exponent of the cutoff.

No time variability has been detected so far from tens of minutes to daily time-scales~\cite{Aharonian:2009zk,Archer:2014jka}. The origin of the emission is still unknown. Compelling scenarios for this VHE emission include : {\it (i)} a cosmic ray source accelerating high energy protons in the vicinity of Sgr A* which produces VHE gamma rays from $\pi_{\rm 0}$ decays originating from cosmic ray protons penetrating the interstellar medium gas~\cite{Aharonian:2006au, Fujita:2015xva}; {\it (ii)} a pulsar wind nebula G359.95-0.04~\cite{Hinton:2006zk}\footnote{With the recent Fermi-LAT detections, the GeV emission from this model significantly underestimates by far the luminosity of the 2FHL J1745.7-2900~\cite{Ackermann:2015uya} and 3FGL J1745.6-2859c~\cite{Acero:2015hja} Fermi sources, given the pronounced peak-like structure exhibited by the IC emission. If the Fermi-LAT and H.E.S.S. emissions are originated by the same mechanism, a pulsar wind nebular scenario may be likely excluded.}; and {\it (iii)} a spike of annihilating DM particles~\cite{2012PhRvD86h3516B}. A template gamma-ray spectrum for the proton-induced emission is a power law with an exponential energy cut-off~\cite{1996SSRv...75..331M}. An alternative scenario which fits the H.E.S.S. spectrum well invokes the annihilation of DM particles to hard channels such as the $\tau^+\tau^-$ and $\mu^+\mu^-$~\cite{2012PhRvD86h3516B}. The DM-induced spectrum template can be interpreted by a power law with super-exponential energy cut off~\cite{Belikov:2013laa}.

Following the latest results from the H.E.S.S. experiment towards the Galactic Center~\cite{Abramowski::2016dhk} and due to the indication that a super-exponential energy cut-off may be a sign of new
physics~\cite{Belikov:2013laa}, we study the potential of observations with future atmospheric Cherenkov arrays such as the Cherenkov Telescope Array (CTA) to unveil the nature of the VHE emission towards the Galactic Center. In this paper, we derive the point estimates of super-exponential fits, revisit the DM interpretation of the Galactic Center VHE gamma-ray emission in light of the new H.E.S.S. data and investigate the potential spectra and morphologies that may be observable by next-generation atmospheric Cherenkov arrays.
We use in this study the CTA observatory~\cite{springerlink:10.1007/s10686-011-9247-0} as an example for future observations of the Galactic Center. We show that higher sensitivity together with improved energy and angular resolutions of CTA can precisely distinguish between exponential and super-exponential models of emission.
Section~\ref{sec:spec} presents the spectral templates together with an update to the DM spectral models for H.E.S.S. data. In section~\ref{sec:mockdata}, we study the potential of CTA to distinguish between the above-mentioned scenarios through spectral measurements.
In section~\ref{sec:discussion} we present our conclusions. In Appendix A we derive the small-angle approximation for the $J$-factor (which parameterises the angular dependence of the expected $\gamma$-ray flux and is usually used in context of annihilating or decaying dark matter) for the power-law-like density distribution. In Appendix B we discuss approximate constraints on the morphology of the central source.

\section{Spectral models for H.E.S.S. data}
\label{sec:spec}
\subsection{Spectral smoothing}
\label{sec:defs}
The differential gamma-ray flux is the number of photons in the direction, specified by polar angle $\psi$ and azimuthal angle $\varphi$, at energy $E_\gamma$ per unit time, unit detection area and unit solid angle is $\frac{dN}{dE_{\gamma} dt \,dA d\Omega}(E_{\gamma},\psi, \varphi)$.
Due to the finite angular and energy resolution of an instrument, the actual observed differential flux
is a convolution of the differential flux of the source with the instrumental angular and energy kernels $K(\psi, \psi', \varphi, \varphi')$ and $M(E, E')$ correspondingly:
\begin{equation}
\frac{\widetilde{dN}}{dt \,dA \, dE_{\gamma} \, d\Omega} (E_{\gamma}, \psi) =
\int \frac{dN}{dt \,dA \, dE_{\gamma} \, d\Omega'}(E_{\gamma},\psi) K(\psi, \psi', \varphi, \varphi') M(E_\gamma, E_\gamma') d\Omega' dE_\gamma'
\label{eqn:flux_conv},
\end{equation}

where $\psi'$ and $\varphi'$ are correspondingly the polar and the azimuthal angles defining the line-of-sight over which kernel $K(\psi, \psi', \varphi, \varphi')$ convolved, $d\Omega' = d\cos\psi' \, d\varphi'$ and $E'$ is the energy over which kernel $M(E,E')$ is convolved.
When integrated over a finite solid angle $\Delta \Omega$ the differential gamma-ray flux from Eq.~\ref{eqn:flux_conv} is called the differential energy spectrum $\Phi(E, \psi, \Delta \Omega)$:

\begin{equation}
\Phi(E, \psi, \Delta \Omega) = \int_{\Delta\Omega}
                         \frac {\widetilde{dN}}
                               {dt \,dA\, dE_{\gamma} \, d\Omega}
                               (E_{\gamma}, \psi) d\Omega.
\label{eqn:big_phi}
\end{equation}

If a physical process responsible for the observed spectrum has no spatial dependence, or it can be neglected, the differential spectrum can be factorized:

\begin{equation}
\frac{dN}{dt \,dA \, dE_{\gamma} \, d\Omega}(E_{\gamma},\psi) = \phi(E_\gamma) J(\psi),
\label{eqn:spec_fact}
\end{equation}
where $J(\psi)$ can be defined as a dimensionless quantity, the differential power of the source per unit solid angle. The power of the source over a finite solid angle will then be
\begin{equation}
J (\psi, \Delta \Omega) = \int \, J(\psi) d\Omega = \int\limits^{2\pi}_0 \int\limits^{\psi}_{\psi+\Delta\psi} J(\psi) \, d\cos\psi \, d\varphi,
\label{eqn:j_omega}
\end{equation}
where $\Delta \Omega = 2\pi(\cos\psi - \cos(\psi + \Delta\psi))$.

The energy smoothing kernel $M(E_\gamma, E_\gamma')$ is taken to be a Gaussian: $C(\sigma(E)) e^{-\frac{(E-E')^2}{2\sigma^2(E)}}$, where $\sigma(E)$ is the energy-dependent dispersion of the instrument and $C$ is the normalization constant.

The convolution of the spectrum with the kernel is the smoothed spectrum $\widetilde{\phi}(E)$, recorded by the instrument:
\begin{equation}
\widetilde{\phi}(E) = \int M(E', E) \phi(E') dE'.
\end{equation}

\subsection{Spectral template}
\label{sec:fits}
In this section we present the parameters of the best-fit functions for $\Phi(E, \Delta\Omega)$, the number of counts per unit area, unit time and unit energy in the central angular region of radius 0.1$^{\circ}$ ($\psi = 0$) and solid angle ($\Delta \Omega_{s} \simeq 10^{-5} \mbox{sr}$), which we denote $(0, \Delta\Omega_s)$.

Following the latest analysis of the Galactic center region by the H.E.S.S. collaboration~\cite{Abramowski::2016dhk} the VHE emission from the annular region between 0.15$^{\circ}$ and 0.45$^{\circ}$
and the respective solid angle of $\Delta \Omega_{\rm ann} \simeq 1.4 \times 10^{-4}\,\mbox{sr}$ can be well fitted with a power law function with
$A_b = (1.92\pm0.08_{\rm stat}\pm 0.28_{\rm syst})\times10^{-12}$ cm$^{-2}$s$^{-1}$TeV$^{-1}$, and $\Gamma_b = 2.32 \pm 0.05_{\rm stat}\pm0.11_{\rm syst}$ \footnote{A section of 66$^{\circ}$ was excluded with the purpose of avoiding contamination from a newly detected source~\cite{Abramowski::2016dhk}.}.

The emission from the central circular region of 0.1$^{\circ}$ radius, namely source HESS J1745-290, is preferably fitted over a pure power law by a power law with an exponential cut-off function (EPL) $\Phi_{EPL} (E) = A_s E^{-\Gamma_s} e^{-E/E_c}$ with $A_s = (2.55 \pm 0.04_{\rm stat}\pm0.37_{\rm syst})\times 10^{-12}$ cm$^{-2}$s$^{-1}$TeV$^{-1}$, $\Gamma_s = 2.14\pm0.02_{\rm stat}\pm 0.10_{\rm syst}$ and $E_c = 10.7 \pm 2.0_{\rm stat}\pm2.1_{\rm syst}$ TeV.
The ratio of the solid angle of the annular region to the solid angle of the central region is $\Delta \Omega_{ann} / \Delta \Omega_{s} = 14.7$ and thus the appropriate normalization of the spectral signal in the central region is $A_b = (1.31\pm0.05_{\rm stat}\pm0.19_{\rm syst})\times10^{-13}\,{\rm cm}^{-2}\rm s^{-1}\rm TeV^{-1}$. While the quality of the fit by a power law with an exponential cut-off to the central source is very good, it is conceivable that the background from the annular region extends into the central region and as such should be treated as fixed component to the source HESS J1745-290, as considered in Ref.~\cite{VianaICRC}. We make an assumption that the spectrum in the annular region is uniform in spherical $\psi, \varphi$ coordinates and therefore can be extended to the central region by appropriately normalizing by the corresponding the solid angle. We note that further studies are required to identify the spatial structure of the diffuse emission as follows from Extender Data Table 1 of Ref.~\cite{Abramowski::2016dhk}.

We compute the best EPL and SEPL fits for the H.E.S.S. pointlike source spectrum of $\Phi(E)$, which is modeled as a superposition of either EPL or SEPL template and a fixed background power-law function as described above. The exponential fit $\Phi_{\rm EPL} (E) = A_s E^{-\Gamma} e^{-E/E_c} $ of H.E.S.S. data with $A_s = (2.41 \pm 0.07) \times 10^{-12} \, \rm cm^{-2}s^{-1}TeV^{-1}$, $\Gamma_c = 2.14 \pm 0.03$ and $\rm E_c = 10.3^{+2.0}_{-1.6} \,TeV$ yields $\chi^2 = 22.9$ for 26 degrees of freedom. Hereafter this function is referred to as template A.
The super-exponential fit $\Phi_{\rm SEPL} (E) = A_s E^{-\Gamma_s} e^{-(E/E_c)^\beta}$ of H.E.S.S. data with $A_s = (2.28 \pm 0.06) \times 10^{-12} \,\rm cm^{-2}s^{-1}TeV^{-1}$, $\Gamma_s = 2.17 \pm 0.03$, $\rm E_c = 11.1 ^{+2.4}_{-1.8} \,TeV$ and $\beta = 1.29^{+0.3}_{-0.21}$ yields $\chi^2 = 22.3$ for 25 degrees of freedom.
If we fix $\beta = 2$ and minimize $\chi^2$ over the rest of the parameters, we get a comparably good fit with $\chi^2 = 24.8$ ($dof = 26$). The parameters are $A_s= (2.16 \pm 0.06)\times 10^{-12} \, \rm cm^{-2}s^{-1}\rm TeV^{-1}$, $\Gamma_s = 2.13 \pm 0.03$ and $E_c = 11.5^{+2.7}_{-2.0}$ TeV. Hereafter this function is referred to as template B.

The best fits are the point estimates of the free parameters of the model given the data: while keeping the parameters associated with the diffuse power-law fixed, we find the global maximum likelihood values of EPL and SEPL templates.
The errors on the point estimates are found by a likelihood ratio test. The test follows a $\chi^2$ distribution with one degree of freedom. The error is found as the difference between the best-fit estimate and the value of the parameter for which $2\Delta \log\mathcal{L}$ is equal to 3.84, which corresponds to 95\% confidence level.

\begin{figure}[hbp]
\centering
\mbox{\includegraphics[width=0.45\linewidth]{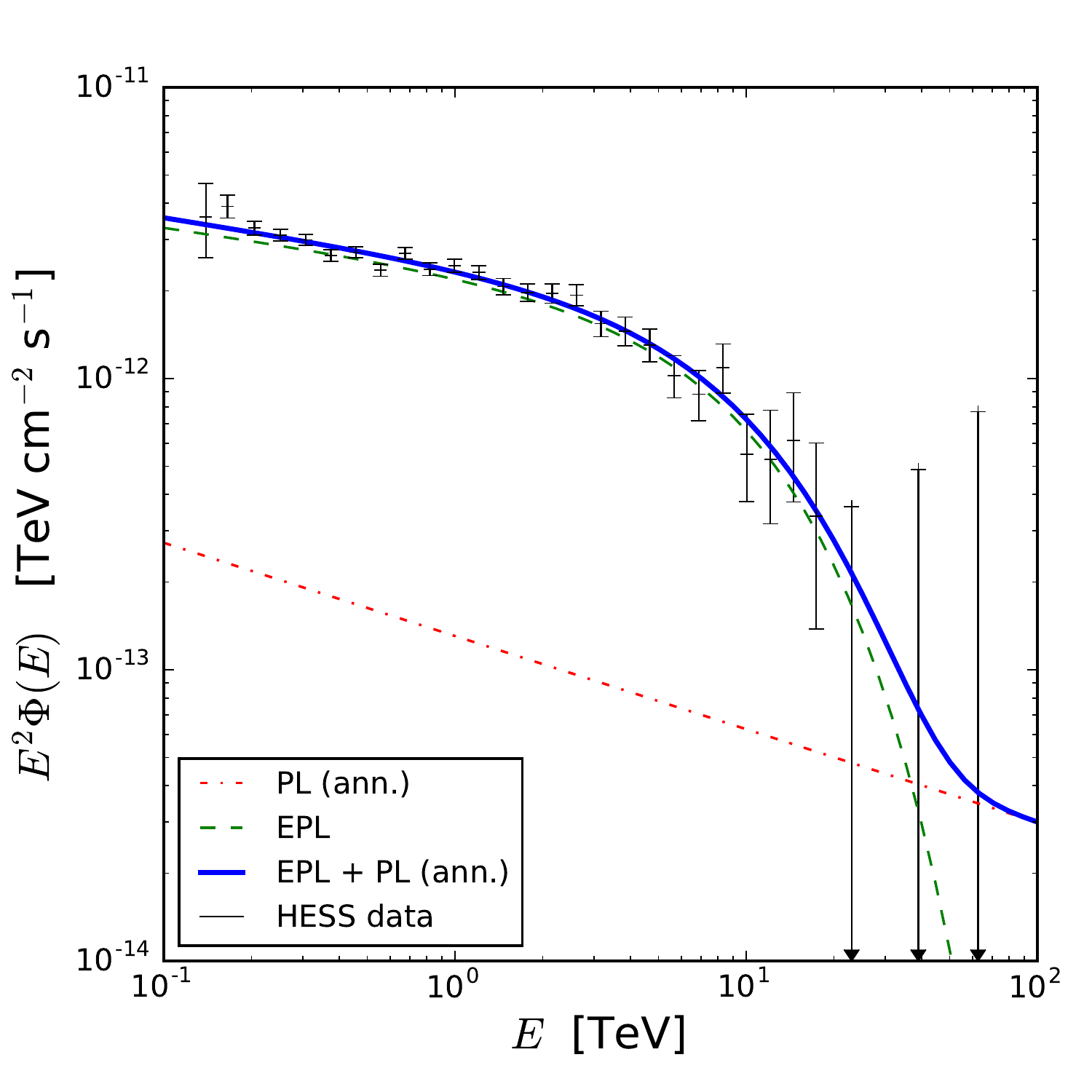}
\includegraphics[width=0.45\linewidth]{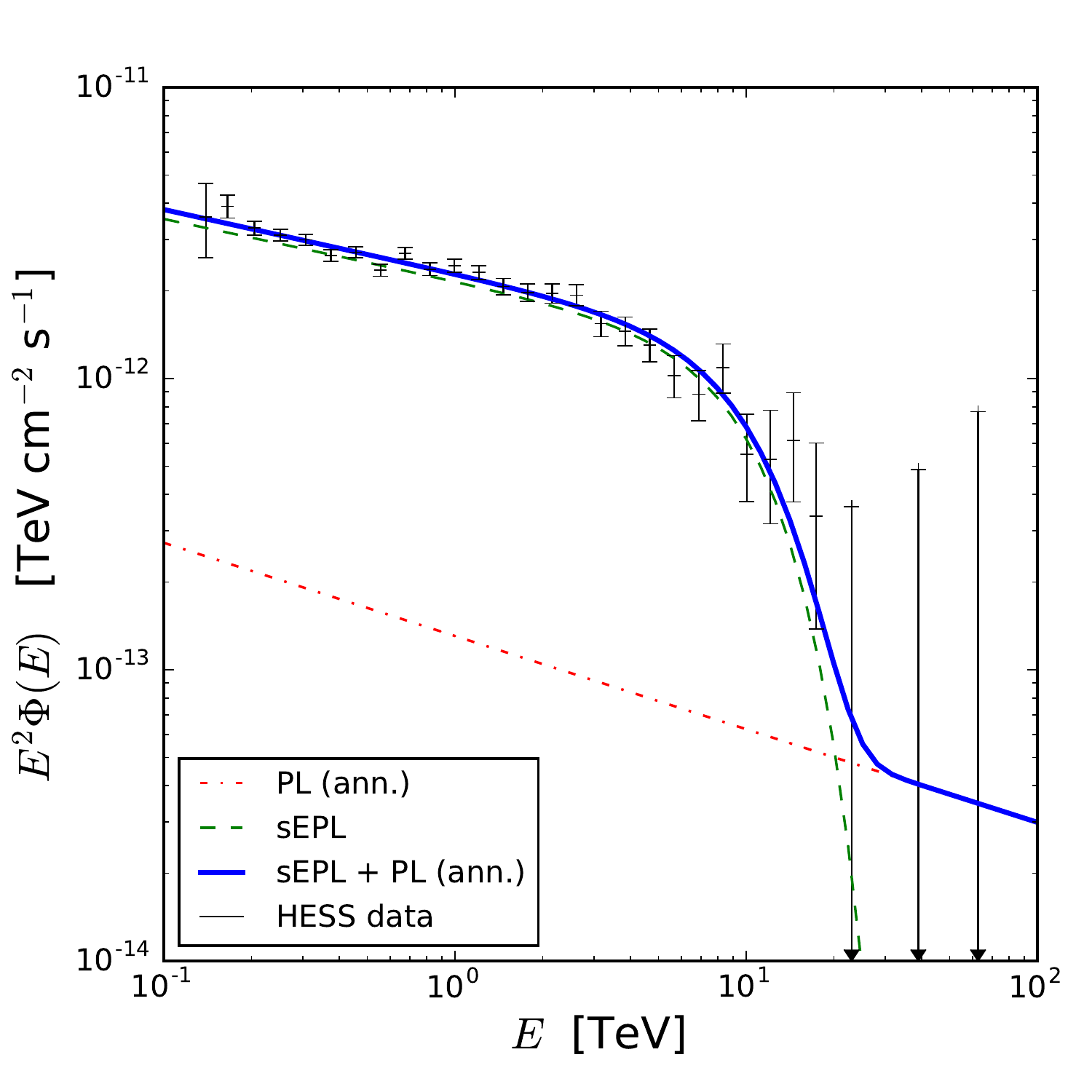}}
\caption{{\it Left panel:} H.E.S.S. spectral data of the Galactic Center, the annulus power-law extrapolation (red, dot-dashed line), the point source exponential cutoff spectrum (green, dashed line), the sum (blue, solid line), $\chi^2/dof = 0.88$. {\it Right panel:} H.E.S.S. data of the Galactic Center, the annulus power-law (red, dot-dashed line), the point source super-exp. cutoff spectrum with $\beta = 2$ (green, dashed line), the sum (blue, solid line), $\chi^2/dof = 0.95$.
}
\label{fig:En2_exp}
\end{figure}

Fig.~\ref{fig:En2_exp} shows the best fit to the data for templates A and B, respectively.
Given the $\chi^2$ values of the two models we conclude that we cannot infer from the H.E.S.S. dataset that the SEPL model is better than the EPL model.

\subsection{Dark matter interpretation}
\label{sec:dm}
The differential gamma-ray flux from self-annihilation of dark matter of mass $m_{\rm DM}$ per unit of time, unit of detection area and unit solid angle $d\Omega = d\cos\psi \, d\varphi$ in the direction $\psi$ and at energy $E_\gamma$ is
\begin{equation}
\frac{dN}{dt \,dA \, dE_{\gamma} \, d\Omega}(E_{\gamma},\psi) = \frac{\langle\sigma v\rangle}{8\pi m^2_{\rm DM}}\frac{dN}{dE_{\gamma}}\left(E_\gamma\right) \int_{los}\rho^2(r(s,\psi))\ ds
\label{eqn:flux}
\end{equation}
where $\langle\sigma v\rangle$ is the velocity-weighted annihilation cross section, $\rho$ is the dark matter mass density integrated over the line of sight (los), and $dN/dE_\gamma = \sum_i R_i \frac{dN_i}{dE_\gamma}$ is the photon spectrum per annihilation summed over the annihilation channels $i$ with branching ratio $R_i$.
We note that the prompt energy spectrum does not have an explicit angular dependence in the case of annihilating dark matter, and can be factorized into an energy-dependent and angle-dependent parts, as in Eq. (\ref{eqn:spec_fact}), with
$J(\psi) = \frac{1}{r_\odot} \frac{1}{\rho^2_0} \int_{los}\rho^2(r(s,\psi))\ ds$ and $\phi(E) = \frac{\langle\sigma v\rangle\rho^2_0\,r_\odot}{8\pi m^2_{\rm DM}} \frac{dN}{dE_\gamma}$, with $r_{\odot} = 8.5 \,\mbox{kpc}$, $\rho_0 = 0.3 \,\mbox{GeV/cm}^3$ and the thermally averaged annihilation cross section $\langle\sigma v\rangle = 3\times 10^{-26}$ cm$^{3}$ s$^{-1}$.

With the definition $\bar J(\psi, \Delta\Omega) = \frac{1}{\Delta \Omega} \int_{\Delta \Omega} J(\psi') K(\psi, \psi', \varphi, \varphi') \,d\Omega' \, \,d\Omega$, Eq. \ref{eqn:big_phi} can be rewritten:

\begin{equation}
\Phi(E_{\gamma}, \psi, \Delta\Omega) \approx \widetilde \phi(E_\gamma) \bar J(\Delta \Omega) \Delta \Omega.
\end{equation}

\begin{figure}[htbp]
\centering
\mbox{
\includegraphics[width=0.5\linewidth,clip=]{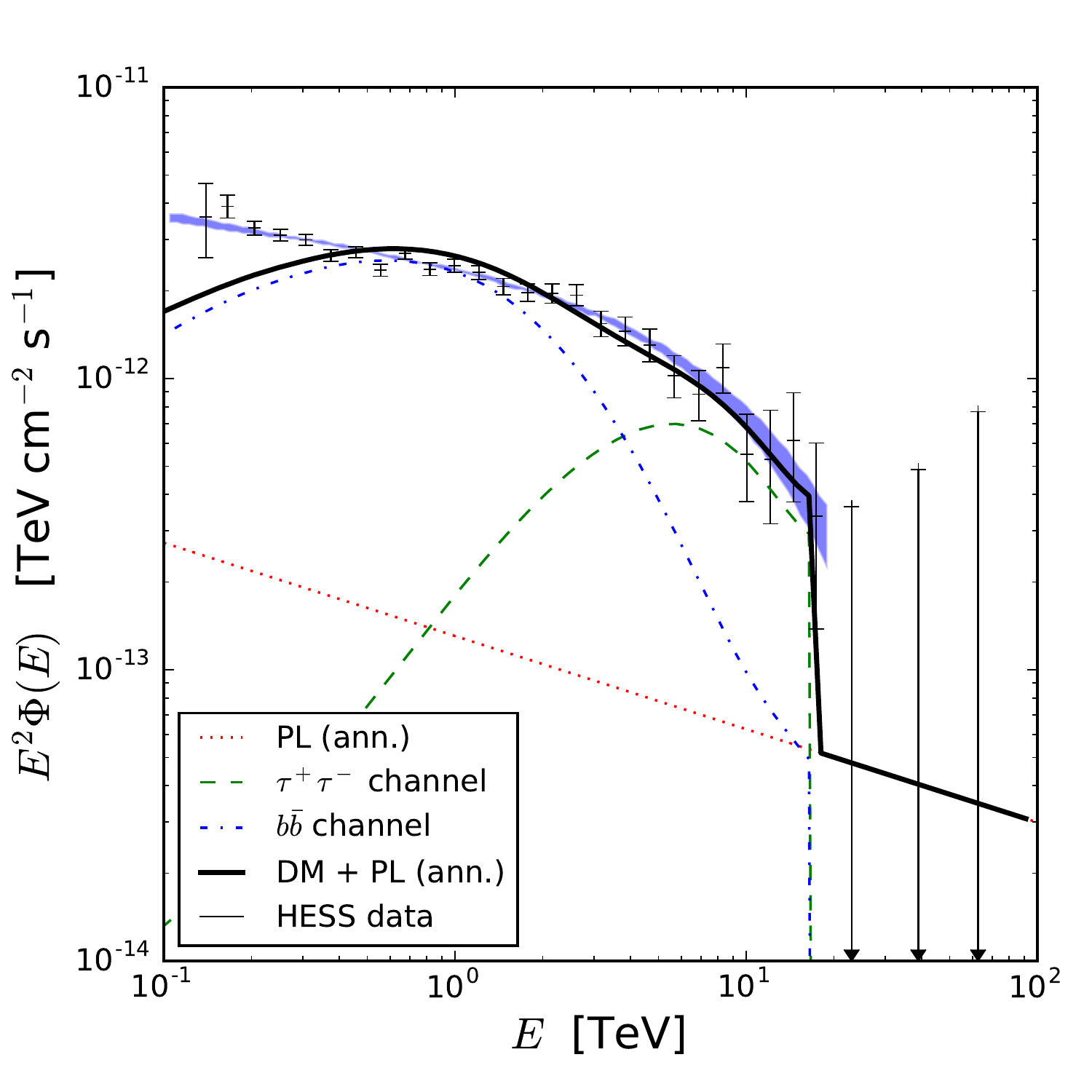}
\includegraphics[width=0.5\linewidth,clip=]{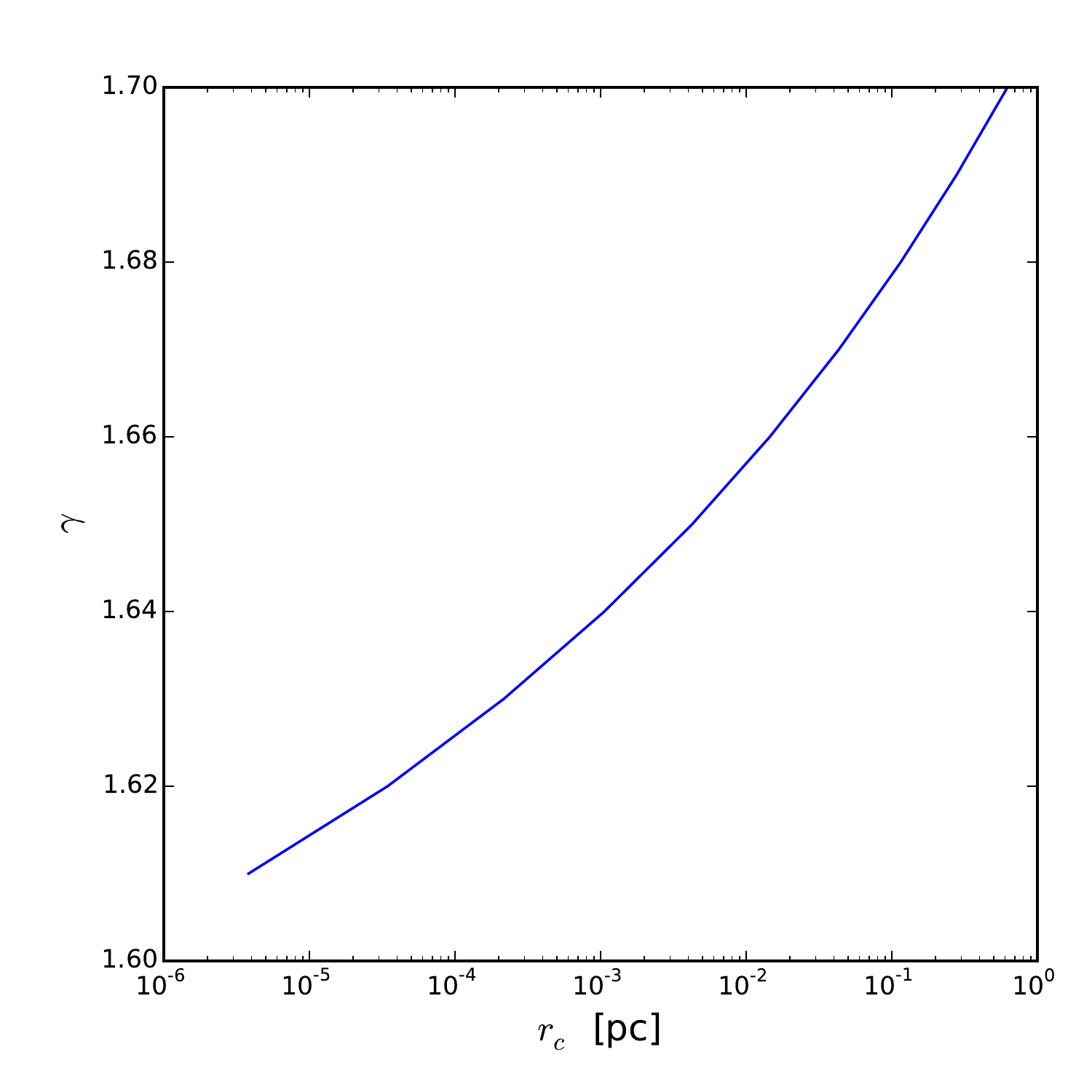}
}
\caption{{\it Left panel:} H.E.S.S. data of the Galactic Center, the annulus power-law (red, dotted line), the contribution of the $\tau^+$ $\tau^-$ channel (green, dashed line) of the gamma-ray spectrum of annihilation dark matter, the contribution of the $b\bar b$ channel (blue, dash-dotted line), and the sum (black, spline line), $\chi^2/dof = 3.4$. {\it Right panel:} The set of values $r_c$-$\gamma$ for which the calculated $J$-factor is equal to best-fit value from the model of the left panel.
}
\label{fig:En2_DM_TABB}
\end{figure}

The best fit model of the pointlike source at the Galactic Center respecting the multi-TeV cut-off (Fig.~\ref{fig:En2_DM_TABB}, left panel) occurs for a combination of $\tau^+ \tau^-$ and $b\bar b$ annihilation channels for the dark matter mass $ m_{DM} = 17.8 \,\rm TeV$,
$\bar J = 1.3 \times 10^5$,
$R_{\tau^+\tau^-} = 0.22$, and $R_{b\bar b} = 1 - R_{\tau^+\tau^-}$.
We note that the dark matter fit respecting the multi-TeV cut-off alone can not explain the H.E.S.S. data below $\sim$300 GeV.
The quality of the fit could be improved if supplemented by an extra component, possibly of pulsar origin.
For instance, a pulsing component above 100 GeV was detected in Crab Pulsar by VERITAS collaboration \cite{Aliu:2011zi}. While most pulsar models predict the spectral cutoff energies in the range of 1-20 GeV, recently theoretical justifications of the pulsar origin of the emission with higher energy cutoff (above 100 GeV), based on inverse Compton effect in the outer gap model context were developed \cite{Aleksic:2011yx, Aharonian:2012zz, Breed:2016wms}. We hypothesize that such a pulsar might be awaiting its detection in the galactic center.

A specific value of $J$-factor places a constraint on the parameters of the dark matter density distribution. In our case the best fit $J$-factor value equal to $1.3 \times 10^5$ is about 390 times greater than that of a power-law like dark matter density profile $\rho \sim r^{-\gamma}$ with $\gamma = 1.5$\footnote{A physical cutoff $r_c = 2r_{\rm Sch}$ is used at small $r$.}.

We note that while the expected from pure N-body cold dark matter simulations slope is close to identity ($\gamma\sim$ 1), it has been proposed that dark matter density undergoes an adiabatic contraction in the vicinity of a super-massive black hole (SMBH)~\cite{Gondolo:1999ef}. The contraction would increase the density in the central region enhancing the dark matter annihilation signal. The amplitude of the enhancement is uncertain \cite{2004PhRvL..93f1302G} and depends, for example, on the growth history of the SMBH \cite{2008PhRvD..78h3506V}. In the Galactic center, the dark matter spike may have been resilient to baryon influence over Gyr time scale~\cite{Lacroix:2015lxa}.

The $J$-factor for the annihilating dark matter derived from a simple power-law-like density profile with $\gamma > 1.5$ would be na\"ively divergent at the origin. However, there exists a physical lower bound, twice the Schwarzschild radius of the Milky Way SMBH, \cite{Sadeghian:2013laa}, below which the dark matter density vanishes.

Furthermore, the dark matter density near the center of the distribution flattens due to self-annihilations \cite{Ahn:2007ty}, and therefore can be modeled by constant density between $2r_{\rm Sch}$ and some critical radius $r_c$. For the Milky Way SMBH $r_{\rm Sch} \simeq 4\times10^{-7}$ pc. The critical radius depends on the spike formation time and can be affected by black-hole mergers \cite{Merritt:2002vj} and in-falling baryonic matter.
We consider a continuous dark matter density template function, which is proportional to a power law $r^{-\gamma}, r \ge r_c$, and a constant for $r < r_c$, and treat $r_c$ as a free parameter along with the exponent of the dark matter profile $\gamma$. We identify the constraints in the $\gamma$-$r_c$ plane, which correspond to the dark matter profile capable of producing the signal of observed magnitude, as shown in Fig.~(\ref{fig:En2_DM_TABB}b). Please note that (a) no $\gamma < 1.61$ can satisfy the $J$-factor constraint ($r_c$ reaches its lowest bound $2r_{\rm Sch}$); (b) for a given $\gamma$-$r_c$ pair, greater $\gamma$ or smaller $r_c$ corresponds to greater $J$-factor and therefore the area above the curve in Fig.~(\ref{fig:En2_DM_TABB}b), is ruled out.

In Appendix B we discuss a derivation of a possible constraint in the $\gamma$-$r_c$ plane from angular extent of the central source.

\section{Generating CTA mock spectral data}
\label{sec:mockdata}
CTA~\cite{cta} will be the next-generation high-energy gamma-ray observatory for ground-based gamma-ray astronomy~\cite{springerlink:10.1007/s10686-011-9247-0}. It will substantially improve the overall performance of the currently operating ground-based Cherenkov telescopes such as H.E.S.S.~\cite{hess}, MAGIC~\cite{magic} or VERITAS~\cite{veritas}. It will operate in the high energy regime from a few tens of GeV to a few hundred TeV, enabling a one-order-of-magnitude increase in sensitivity in the TeV regime  and a factor of two-to-three improvement in angular resolution compared to current instruments, achieving to better than 0.03$^{\circ}$ above 1 TeV~\cite{springerlink:10.1007/s10686-011-9247-0}.
This observatory will be composed of two sites, one in each hemisphere, allowing for full sky coverage.
The exact layouts of the two arrays are not yet settled and extensive Monte Carlo simulations have been performed to characterize different candidate arrays in terms of instrument response function, i.e. effective area, angular and energy resolution and background rejection capabilities. We study the spectral and morphological prospects of CTA using the instrument response functions of the candidate array for the Southern array given in Ref.~\cite{ctaIRF}. In particular we use the models for the background rate, the collection area and the energy resolution data of the Southern site which is preferred for Galactic center observations.

Using templates A and B with an exponential cutoff and a super exponential cutoff for the spectral shape of the signal respectively from section \ref{sec:fits}, the power law shape for the astrophysical background and the instrumental noise we generate the mock data. The mean number of events in the bin ($i$, $j$) at energy $(E_{\gamma i},\Delta E_{\gamma i})$  and angle $j$ $(\psi_j,\Delta \Omega_j)$ registered by the telescope array with an effective collecting area $A_{\rm eff}(E)$ during time $\Delta t$, given the flux $\Phi(E, \psi_j, \Delta\Omega_j)$, defined in Eq.~\ref{eqn:big_phi}, can be estimated as
\begin{equation}
N_{ij} = \int_{\Delta t} dt \int_{total\,area} dA \, A_{\rm eff} (E) \int_{\Delta E_{\gamma i}}dE_\gamma \,\Phi(E_\gamma, \psi_j, \Delta \Omega_j),
\label{eqn:counts}
\end{equation}
where $\Delta E_{\gamma i}$ is the width of the energy bin $i$ centered at $E_{\gamma i}$ and $\Delta \Omega_j$ denotes the solid angle bin $j$ centered at $\psi_j$.
The energy smoothing kernel is a Gaussian with the variance of energy $\sigma_i = E_i \delta_{\rm res} / \sqrt{8\ln 2}$, related to energy resolution of the instrument $\delta_{\rm res} = \Delta E_i/ E_i$ and $C(\sigma_i) = \frac{1}{\sigma_i\sqrt{2\pi}}$.

Since in our case the normalization constants correspond to the central angular bin on $\Delta \Omega \simeq 10^{-5}$ sr, we omit the $j$ index and calculate the mean values of the counts for different energy bins for the signal $N^s_i$, annular power law background $N^a_i$ and the instrumental background $N^a_i = \frac{\Delta t}{\Delta E_i} \int_{\Delta E_i} R(E) dE$, where $R(E)$ [\mbox{s}$^{-1}$] is the instrumental background rate. Once we have the mean number of events for a given energy bin, we generate $N_r = 1000$ realizations of each the processes   and calculate the statistical error assuming a Poisson distribution. We assume the systematic error $\delta_{syst}$ to be 1\% below the cutoff energy of 5 TeV and 10\% above. The total error includes the statistical as well as instrument specific systematic error $\delta_{\rm syst}$. We take the energy bins with logarithmic spacing and with 5 bins per decade from from 100 GeV to 100 TeV.

\begin{figure}[H]
\centering
\mbox{\includegraphics[width=0.5\linewidth,clip=]{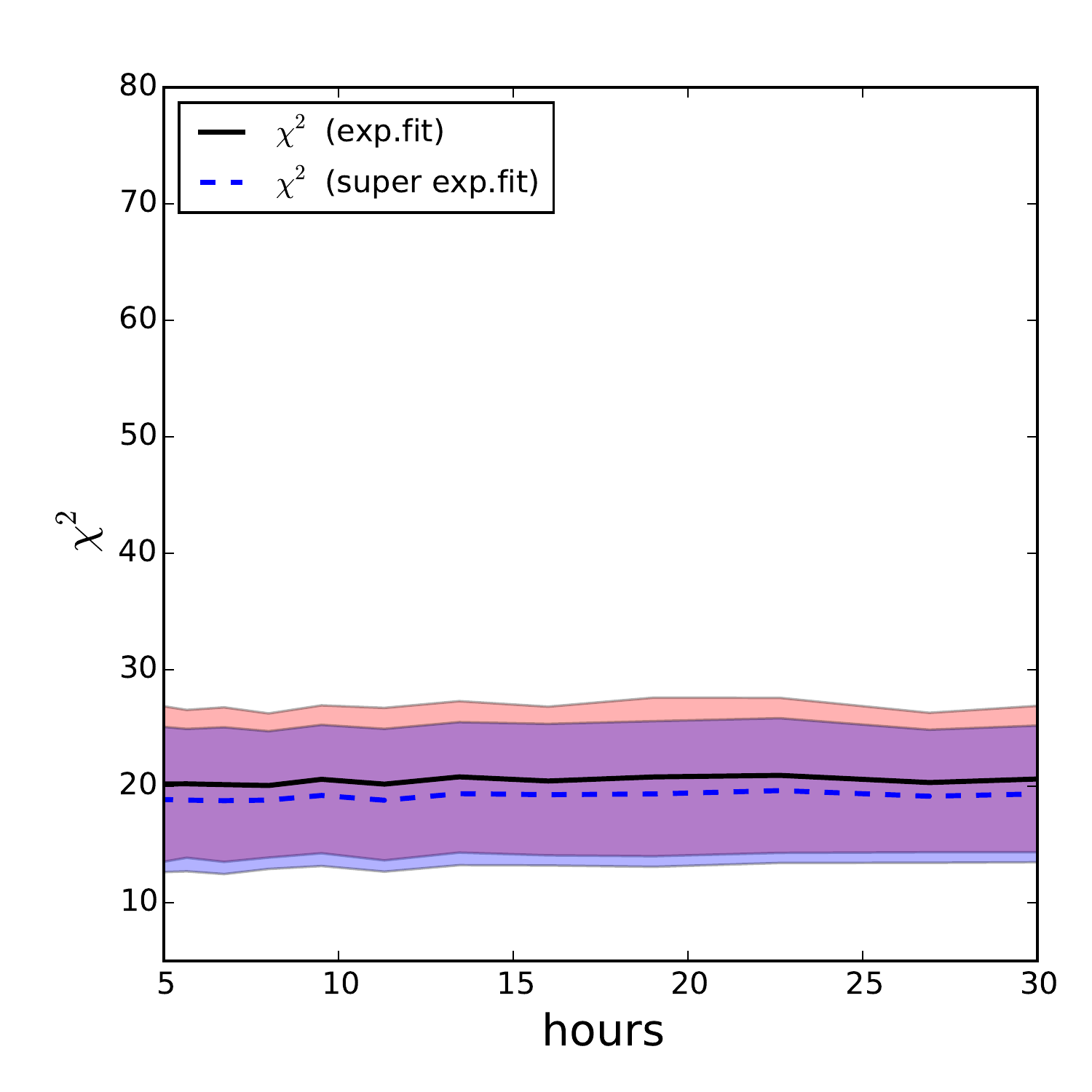}
\includegraphics[width=0.5\linewidth,clip=]{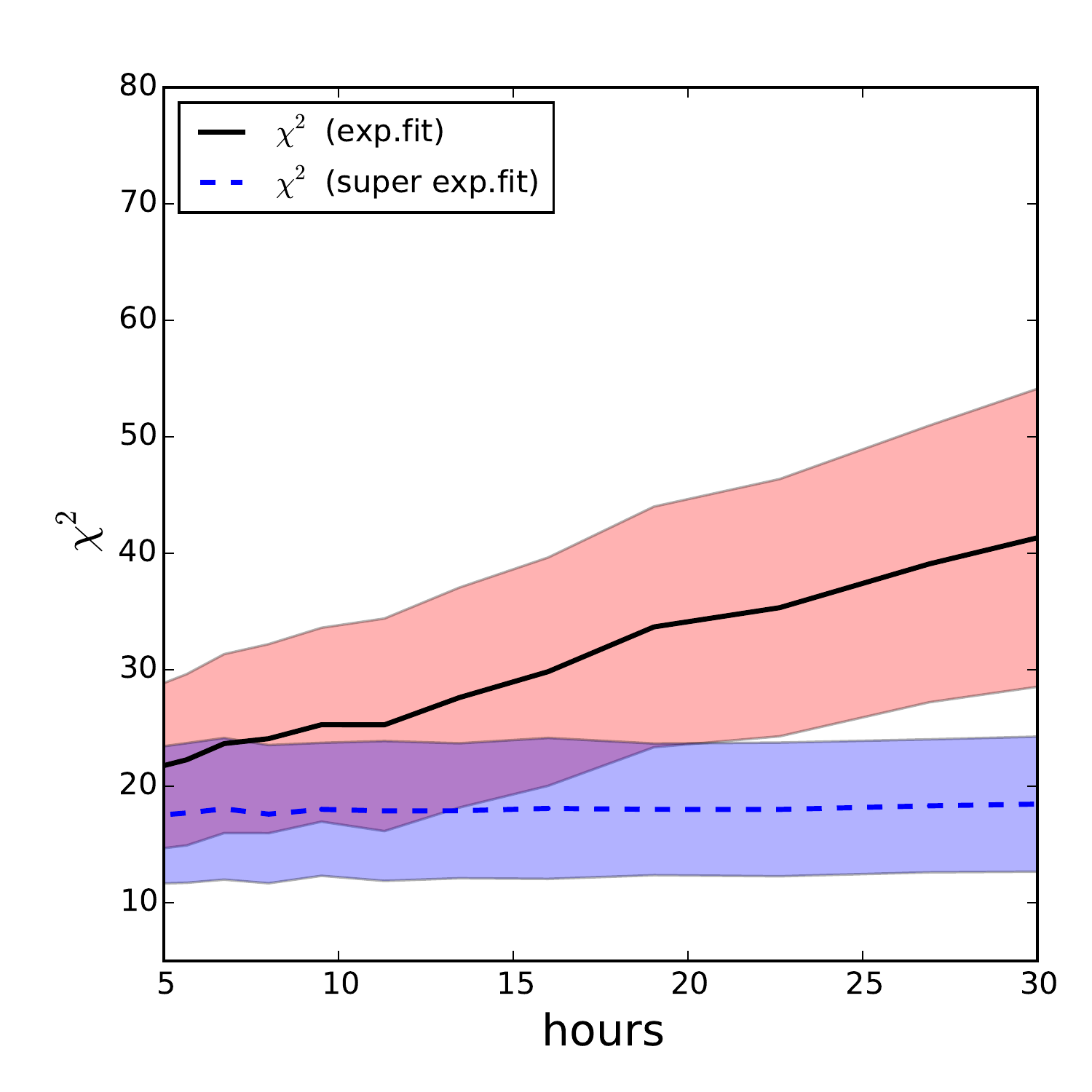}}
\caption{Left: $\chi^2$ for the exponential and super-exponential fits to mock data generated with the exponential template A. Right: $\chi^2$ for the exponential and super-exponential fits to mock data generated with the super-exponential template B. In both panels black solid lines denote the super-exponential fits, while blue dashed lines denote the exponential fits. The red and the blue shaded regions denote one standard deviation regions for the exponential and the super-exponential fits.
}
\label{fig:chi2_expSignal}
\end{figure}

\begin{figure}[ht]
\centering
\mbox{\includegraphics[width=0.5\linewidth,clip=]{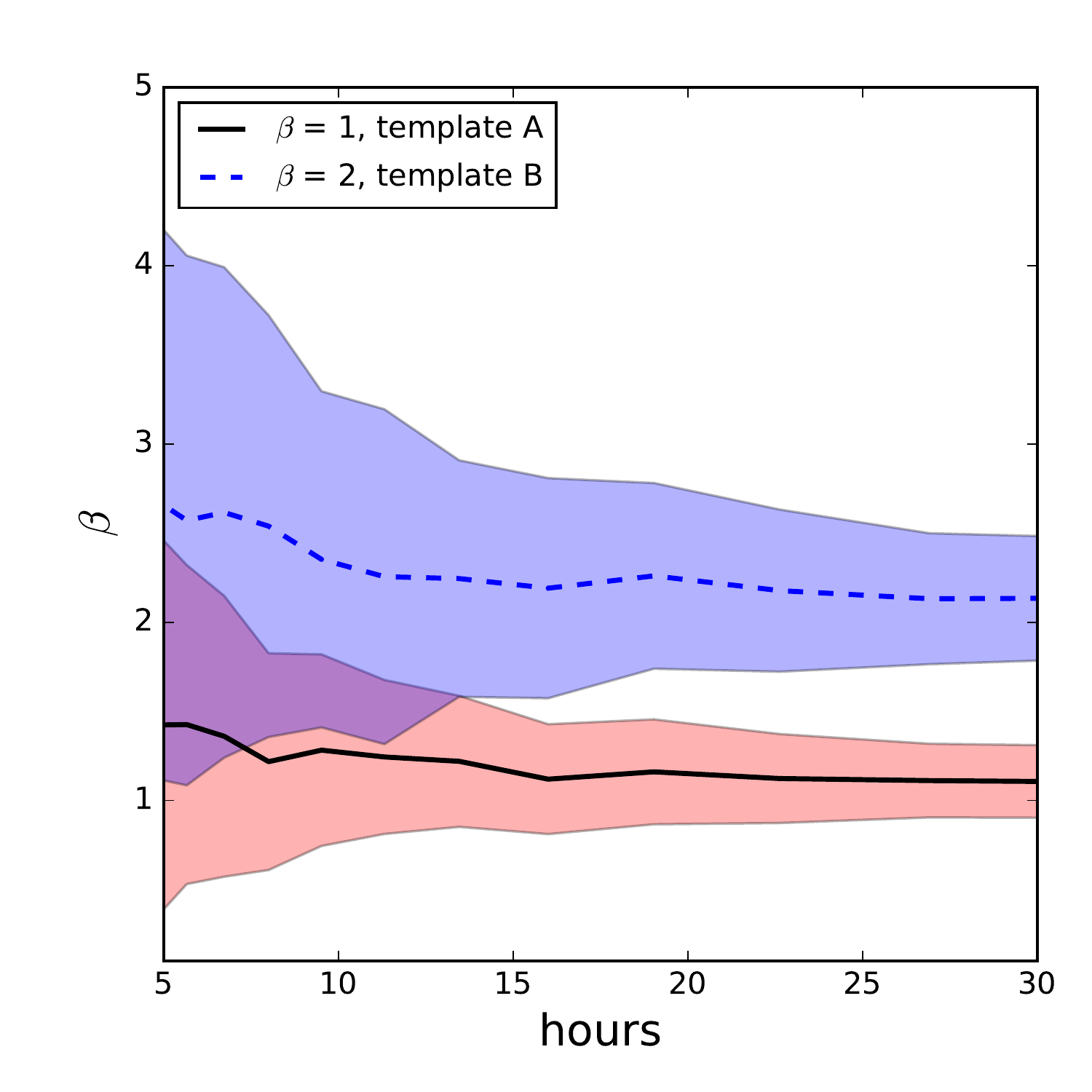}
}
\caption{$\beta$ of the super-exponential fits to mock data generated with the exponential template A and
$\beta$ of the super-exponential fits to mock data generated with the exponential template B. The black solid line denotes the super-exponential fits to template A, while the blue dashed line denotes the super-exponential fits to template B. The red and the blue shaded regions denote one standard deviation regions for the fits to template A and template B correspondingly.}
\label{fig:beta_expSignal}
\end{figure}

Each spectrum is then fitted with the exponential and the super-exponential template fitting functions. The means and the standard deviations of $\chi^2$ and $\beta$ (super-exponential only) are calculated and presented in Figs.~\ref{fig:chi2_expSignal} and \ref{fig:beta_expSignal}. The left panel of Fig.~\ref{fig:chi2_expSignal} confirms that in case the true signal is just an exponential cutoff, there will be no gain in fitting it with a super-exponential function. The right panel of Fig.~\ref{fig:chi2_expSignal} shows that CTA will be able to reliably distinguish a SEPL from EPL models in about 20 hours of observation time.
The two template models are compared via a likelihood ratio test. The test follows a $\chi^2$ distribution with one degree of freedom. Template B model fitted to the data is considered significantly better that template A model for $2\Delta \log\mathcal{L}$ greater than 3.84, which corresponds to 95\% confidence level. This occurs for the observation time of approximately 5 hours.

In Fig. \ref{fig:beta_expSignal} the best-fit estimates of parameter $\beta$ to templates A and B are presented as a function of observation time. We confirm that for template A the estimate of $\beta$ converges to 1, while for template B the estimate of $\beta$ converges to 2, as expected.

\section{Discussion}
\label{sec:discussion}
The observations by H.E.S.S. of the Galactic Center region confirm a VHE source coincident with the Galactic Center up to the resolution of H.E.S.S. Its energy spectrum is well described by a power law with an exponential cutoff function, while its angular spectrum is pointlike.

In this paper we verify that it can be as well represented by a power law with an super-exponential cutoff function. We find that TeV cutoff in the energy spectrum of the central source can be fit by a dark matter model, but overall the dark matter fit cannot compete with the exponential cutoff models in the whole range of energies.
From the normalization of the source we rule out dark matter density profiles with slopes $\gamma \le 1.5$. By invoking the cutoff radius, below which the dark matter remains constant, we identify the constraints in $\gamma-r_c$ plane, based on the magnitude of the flux.

We also evaluate the future prospects for the CTA experiment to distinguish the exponential cutoff against the super-exponential cutoff energy spectra and find that with 10-20 hours of observation one would be able to choose between the exponential cutoff against the super-exponential cutoff templates of the description of the central source.

In Appendix A we confirm that a power-law-like density profile leads to a power-law-like $J$-factor in the small angle approximation. In Appendix B we propose a method of deriving approximate constraints on the morphology of the central source.

\section{Acknowledgments}
This research has been supported at IAP by the ERC project 267117 (DARK) hosted by Universit\'e Pierre et Marie Curie - Paris 6.
We thank an anonymous referee for useful comments and suggestions.

\section{Appendix A}
\label{app:a}
The angular spectrum of prompt radiation from annihilating dark matter is $J(\psi) = \frac{1}{r_\odot\rho^2_0} \int\limits^\infty_{-r_\odot} \rho^2(r) ds,$
where $r = \sqrt{s^2 + r_\odot^2 - 2sr_\odot\cos{\psi}}$. At small $r$ the density distribution (if cuspy) can be represented by a power law:
\begin{equation}
\rho(r) = \rho' \left(\frac{r}{a}\right)^{-\gamma}
\end{equation}

We note that $\rho' = \frac{3 - \gamma}{3} \rho_0 \Delta_V c^\gamma$ and $a^3 = \frac{M}{\frac{4}{3} \pi \Delta_V(0) \rho_0 c^3}$, where $\Delta_V \approx 340$ is the virial overdensity,
$c$ is the halo concentration parameter and $M$ is the halo mass \cite{Bergstrom:1997fj}.

Substituting of the power law yields
\begin{align}
J(\psi) = \left(\frac{\rho'}{\rho_0}\right)^2 \left(\frac{a}{r_\odot}\right)^{2\gamma} \int\limits^\infty_{-1} (1+x^2 - 2 x \cos\psi)^{-\gamma} dx,
\end{align}
with $x = s/r_\odot$. One can show that
\begin{align}
\label{eq:hyp}
\int\limits^\infty_{-1} (1+x^2 - 2 x \cos\psi)^{-\gamma} dx = \frac{\Gamma(1/2)\Gamma(\gamma-1/2)}{\Gamma(\gamma)} + 2\tan \frac{\psi}{2}\, _2F_1(\gamma, 1/2; 3/2; -\tan^2 \frac{\psi}{2}).
\end{align}

Keeping the first term in the expansion of the second term of Eq.~(\ref{eq:hyp}) we obtain
\begin{align}
J(\psi) \simeq (\sin\psi)^{1-2\gamma} \left(\frac{\rho'}{\rho_0}\right)^2 \left(\frac{a}{r_\odot}\right)^{2\gamma} \left (\frac{\Gamma(1/2)\Gamma(\gamma-1/2)}{2\Gamma(\gamma)} + \tan \frac{\psi}{2} \right).
\end{align}

For $\psi \ll 1$ and $J$-factor defined as $J(\Delta \Omega) = 2\pi \int\limits^{\psi_1}_{\psi_2} J(\psi) d\cos\psi$ the following holds
\begin{align}
J(\Delta \Omega) = 2\pi \left. \left(\frac{\rho'}{\rho_0}\right)^2 \left(\frac{a}{r_\odot}\right)^{2\gamma} \psi^{3-2\gamma}  \left ( \frac{2}{3-2\gamma}  \frac{\Gamma(1/2)\Gamma(\gamma-1/2)}{2\Gamma(\gamma)}   + \frac{1}{4-2\gamma} \psi + o(\psi) \right ) \right |^{\psi_2}_{\psi_1}.
\label{jfactfinal}
\end{align}

While we were in final stages of preparation of this manuscript, Ref. \cite{PhysRevD.93.103512} was published in PRD. Considering the difference in notation and different integration limits used in our papers, Eq. \label{jfactfinal} of current manuscript is in agreement with Eq. (12) of the aforementioned publication.

\section{Appendix B: Constraining the morphology of J1745-290}
\label{app:b}
In this appendix we sketch a derivation of approximate constraints on the parameters of a density profile of the signal using the background errors. In this section we assume that the differential spectrum can be factorized into an energy dependent and an angular part \ref{eqn:spec_fact}.
While the overall magnitude of the signal from the pointlike source HESS J1745-290 allowed us to derive constraints on the parameters defining the angular distribution of the Fig.~\ref{fig:En2_DM_TABB}(b), here we explore the possibility of constraining the angular shape of the central source by combining the errors on the background energy spectrum and the spectral normalization of the source.

We recount that the spectral fits from Section~\ref{sec:fits} are made for the number of counts per unit area, unit time and unit energy in the central angular bin $(0, \Delta\Omega_0)$ and have the functional form $A_s h_s(E)$ for the signal and $A_b h_b(E)$ for the background.
The observed number of counts is smoothed and integrated over the central angular bin $(0, \Delta\Omega_0)$ and averaged over the energies of the differential spectrum, as in Eq.~(\ref{eqn:spec_fact}):

\begin{equation}
\Phi(E, 0, \Omega_0) = \tilde J(0, \Delta \Omega_0)
\phi(E)
\end{equation}

We rewrite the number of counts in a solid angle $(\psi, \Delta \Omega_1)$ as

\begin{equation}
\Phi(E, \psi, \Omega_1) = \Phi(E, 0, \Omega_0) \frac{\tilde J(\psi, \Delta \Omega_1)}{\tilde J(0, \Delta \Omega_0)}
\end{equation}

We argue that since HESS J1745-290 is identified as a pointlike source, the number of counts due to the pointlike source outside of the central angular region should be less than the error on the number of counts (Eq. \ref{eqn:flux_conv}) from the background for a finite energy interval $\Delta E$ : $N_s (\psi, \Delta \Omega_1, \Delta E) \lesssim \delta N_b (\psi, \Delta \Omega_1, \Delta E)$, where
\begin{equation}
N_s (\psi, \Delta \Omega_1, \Delta E) = \tilde J_s(\psi, \Delta \Omega_1) \int \tilde \phi_s(E) dE =
\frac{\tilde J_s(\psi, \Delta \Omega_1)}{\tilde J_s(0, \Delta \Omega_0)} \int A_s h_s(E) dE,
\end{equation}
and similarly

\begin{equation}
\delta N_b (\psi, \Delta \Omega_1, \Delta E) = \tilde J_b(\psi, \Delta \Omega_1) \int \delta \tilde \phi_b(E) dE =
\frac{\tilde J_b(\psi, \Delta \Omega_1)}{\tilde J_b(0, \Delta \Omega_0)}
\int \delta (A_b h_b(E)) dE.
\end{equation}

Combining the previous three equations we obtain:
\begin{equation}
g(\psi) = \frac{\tilde J_s(\psi, \Delta \Omega_1)}{\tilde J_s(0, \Delta \Omega_0)}
\frac{\tilde J_b(0, \Delta \Omega_0)}{\tilde J_b(\psi, \Delta \Omega_1)}  \lesssim \frac{\int \delta (A_b h_b(E)) dE}{\int A_s h_s(E) dE}.
\label{eqn:ineq}
\end{equation}

Since the background energy spectrum is a power-law $h_b(E) = E^{-\Gamma}$, the error is $\delta (A_b h_b(E)) =  A_b h_b(E)\sqrt{\left(\frac{\delta A_b}{A_b}\right)^2 + (\delta \Gamma \ln E)^2}$.

Inequality (\ref{eqn:ineq}) should hold for all values $E$, $\Delta E$, $\psi$ and $\Delta\psi$. By minimizing the right hand side over possible $E$ and $\Delta E$ and maximizing the left hand side over possible $\psi$ and $\Delta\psi$ we would be able to identify the critical parameters of $J(\psi)$ for which the inequality is saturated.

In our case $h_b = E^{-\Gamma_b}$, $h_s = E^{-\Gamma_s} e^{-E/E_c}$ and the infimum of the ratio of the integrals is reached at the same time as the infimum of the ratio $h_b/h_s$. If we neglect the functional dependence on $\Gamma$ and consider that $\delta (A_b h_b(E)) = h_b(E) \delta A_b$, the solution can be found explicitly: $E_m =  (\Gamma_s - \Gamma_b)E_c$.
In case we respect the functional dependence $\Gamma$, we numerically find that for the best-fit parameters  and their errors derived in Sec. \ref{sec:fits} :
\begin{equation}
\inf \frac{\int \delta (A_b h_b(E)) dE}{\int A_s h_s(E) dE} \approx 0.01\,.
\label{eqn:inf}
\end{equation}

The left hand side of Eq. (\ref{eqn:ineq}) bears only angular dependence with $\tilde J_b(\psi, \Delta \Omega) = 2\pi(\cos\psi - \cos(\psi + \Delta\psi))$ and
\begin{equation}
\tilde J_s(\psi)
= \frac{1}{2\pi\sigma^2} \int \, d\alpha \, d\rho \, \rho \, e^{-\rho^2/2\sigma^2} J_s((\psi^2 + \rho^2 - 2\psi\rho\cos\alpha)^{1/2}),
\label{eqn:j_conv}
\end{equation}
where $\sigma \simeq 0.1^{\circ}$ is the angular resolution of H.E.S.S.

Motivated by sections \ref{sec:dm} and \ref{app:a} we could represent $J(\psi)$ as a power-law with a cutoff at $\psi_c$.
We note that the 2D convolution Eq.(\ref{eqn:j_conv}) can display two different behaviors depending on the ratio of parameters $\psi_c$ and $\sigma$. If $\psi_c \ll \sigma$, $J(\psi)$ is effectively a delta function $\delta(\psi)$ and the convolution approximately yields a Gaussian. If $\sigma \ll \psi_c$, the Gaussian kernel appears to be a singular normalizable function and the convolution approximately yields $J(\psi)$.
The numerical estimates we carried out show that

\begin{equation}
    \frac{\tilde J_s(\psi, \Delta \Omega_1)}{\tilde J_s(0, \Delta \Omega_0)}
    \frac{\tilde J_b(0, \Delta \Omega_0)}{\tilde J_b(\psi, \Delta \Omega_1)} \gg 0.01,
\end{equation}

in the range $r_c < 1$ and $1.5 < \gamma < 2.25$. We conclude that the proposed sketch of deriving constraints does not prove to be useful given the current estimates of the parameters of the pointlike source and the background. For analysis of the raw data containing recorded events, one could use more sophisticated Bayesian methods, such as inference on mixtures of distributions, and derive probabilistically interpretable constraints.

\bibliography{bibl}

\end{document}